\documentclass[pra,aps,onecolumn,superscriptaddress]{revtex4-2}

\usepackage[hyperindex,breaklinks]{hyperref}
\usepackage{CJK}
\usepackage{graphicx}
\usepackage{subcaption}

\usepackage{mathptmx}
\usepackage{amsmath}
\usepackage{amsfonts}
\usepackage{amssymb}
\usepackage{mathrsfs}
\usepackage{float}
\usepackage{esint}
\usepackage{xcolor}

\usepackage{braket}
\usepackage[left=1in, right=1in, top=1in, bottom=1in]{geometry}

\usepackage{tikz}
\usetikzlibrary{shapes.geometric, arrows}
\usepackage{qcircuit}

\makeatletter
\long\def\@makecaption#1#2{%
  \vskip\abovecaptionskip
  \sbox\@tempboxa{#1: #2}%
  \ifdim \wd\@tempboxa >\hsize
    #1: #2\par
  \else
    \global \@minipagefalse
    \hb@xt@\hsize{\hfil\box\@tempboxa\hfil}%
  \fi
  \vskip\belowcaptionskip}
\makeatother

\def\be{\begin{equation}}
\def\ee{\end{equation}}
\def\ba{\begin{eqnarray}}
\def\ea{\end{eqnarray}}

\def\pnp{${\text P}^{\sharp \text{P}}$ }

\begin{document}
\begin{CJK*}{UTF8}{gbsn}
\title
{The Power of Lorentz Quantum Computer}
\author{Qi Zhang(张起)}
\affiliation{College of Science, Liaoning Petrochemical University,
Fushun 113001, China}

\affiliation{Liaoning Provincial Key Laboratory of Novel Micro-Nano Functional Materials,
Fushun 113001, China}

\author{Biao Wu(吴飙)}
\affiliation{International Center for Quantum Materials, Peking University, 100871, Beijing, China}
\affiliation{Wilczek Quantum Center, School of Physics and Astronomy, Shanghai Jiao Tong University, Shanghai 200240, China}

\begin{abstract}
We demonstrate the superior capabilities of the recently proposed Lorentz quantum computer (LQC) compared to conventional quantum computers. We introduce an associated computational complexity class termed bounded-error Lorentz quantum polynomial-time (BLQP), demonstrating its equivalence to the complexity class ${\text P}^{\sharp \text{P}}$. We present LQC algorithms that efficiently solve the problem of maximum independent set, PP (probabilistic polynomial-time), and consequently ${\text P}^{\sharp \text{P}}$, all within polynomial time. Additionally, we show that the quantum computing with postselection proposed by Aaronson can be efficiently simulated by LQC, but not vice versa.
\end{abstract}

\maketitle

\section{Introduction}

Theoretical computing models are fundamentally important in computer science, shaping our understanding of the core principles, boundaries, and possibilities of computing~\cite{Arora,Papadimitriou}. Models like the Turing machine and the quantum Turing machine are physically plausible, serving as abstractions of real-world computers. Conversely, some models are not physically realizable but remain crucial for exploring and clarifying computing problem complexities. A prime example is the non-deterministic Turing machine (NDTM), which, despite its theoretical nature, is extensively utilized in the analysis of complexity classes. In particular, the complexity class NP is alternatively defined as a set of languages decidable by an NDTM within polynomial time.

Quantum computer with postselection is another theoretical model that is not physically sound
because ``the ability to postselect on a measurement yielding a specific outcome" is beyond
the basic principle of quantum mechanics~\cite{knill,Post}. Nevertheless, this model is theoretically valuable, illuminating the complexity class PP (probabilistic polynomial-time) and uncovering connections between quantum mechanics' core principles and the constraints of quantum computing~\cite{Post}.

The theoretical framework for the Lorentz Quantum Computer (LQC) has recently been introduced~\cite{LC}, featuring the innovative concept of the hyperbolic bit (hybit), which evolves via complex Lorentz transformations. This framework draws inspiration from the dynamics of bosonic Bogoliubov quasi-particles, with significant references including works such as~\cite{wunjp,Zhang}. Despite its theoretical appeal, the current model lacks practical feasibility.

The concept of an indefinite inner product, originally proposed by Dirac, has been further developed to tackle convergence issues in quantized field theories. This concept is integral to the $\lambda$-limiting process, which considers the representation of physical observables through self-adjoint operators, as opposed to Hermitian ones~\cite{Dirac,Pauli}. There is optimism that the practicability of this model will enhance as future research achieves a unified theory that incorporates both quantum mechanics and gravity under the Lorentz quantum mechanics framework.

Despite its current constraints, the LQC model remains a subject of substantial theoretical interest. It has been highlighted in~\cite{LC} that LQC potentially offers exponential acceleration in algorithms such as the Grover search algorithm~\cite{Grover}, surpassing the capabilities of conventional quantum computers.

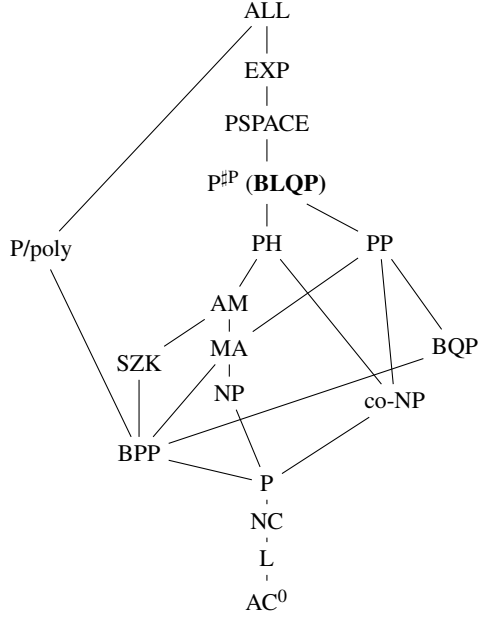
\begin{figure}[t]
\centering
\begin{tikzpicture}
 \tikzstyle{norm}=[rectangle, align=center]

      \node(all){ALL};
      \node(exp)[norm,below of=all,yshift=2mm]{EXP};
      \node(pspace)[norm,below of=exp,yshift=3mm]{PSPACE};
      \node(pnp)[norm,below of=pspace,yshift=2mm]{P$^{\sharp\text{P}}$ (\bf BLQP)};
      \node(ph)[norm,below of=pnp,yshift=2mm]{PH};
      \node(am)[norm,below of=ph,xshift=-5mm,yshift=2mm]{AM};
      \node(poly)[norm,below of=ph,xshift=-3cm,yshift=9mm]{P/poly};
      \node(ma)[norm,below of=am,yshift=4mm]{MA};
      \node(np)[norm,below of=ma,yshift=4mm]{NP};
      \node(szk)[norm,below of=am,xshift=-12mm,yshift=2.0mm]{SZK};
      \node(bpp)[norm,below of=szk,yshift=-2mm]{BPP};
      \node(p)[norm,below of=np,xshift=5mm,yshift=-2mm]{P};
      \node(nc)[norm,below of=p,yshift=5mm]{NC};
      \node(l)[norm,below of=nc,yshift=5mm]{L};
       \node(ac0)[norm,below of=l,yshift=4.5mm]{AC$^0$};
       \node(pp)[norm,right of=ph,xshift=5mm]{PP};
       \node(conp)[norm,right of=np,xshift=12mm,yshift=-1mm]{co-NP};
      \node(bqp)[norm,right of=am,xshift=20mm,yshift=-6mm]{BQP};
      \draw (all) -- (exp);
      \draw (all) -- (poly);
      \draw (exp) -- (pspace);
      \draw (pspace) -- (pnp);
      \draw (pnp) -- (ph);
      \draw (ph) -- (am);
      \draw (am) -- (ma);
      \draw (ma) -- (np);
      \draw (am) -- (szk);
     \draw (szk) -- (bpp);
     \draw (poly) -- (bpp);
     \draw (ma) -- (bpp);
     \draw (np) -- (p);
     \draw (bpp) -- (p);
     \draw (p) -- (nc);
       \draw (l) -- (nc);
       \draw (l) -- (ac0);
       \draw (pnp) -- (pp);
        \draw (pp) -- (bqp);
       \draw (pp) -- (ma);
         \draw (pp) -- (conp);
         \draw (ph) -- (conp);
         \draw (bqp) -- (bpp);
           \draw (conp) -- (p);
   \end{tikzpicture}

   \caption{The hierarchy diagram for major complexity classes.
   For two connecting classes, the class below is  included within the class above.
   BLQP is a complexity class defined for Lorentz quantum computer in parallel to BQP for conventional
   quantum computer.   This diagram
   without BLQP can be found at www.complexityzoo.com.}
   \label{t1}
\end{figure}

In this study, we systematically explore the capabilities of LQC. By analogy with the BQP (bounded-error quantum polynomial-time) complexity class for quantum computing~\cite{Nielson}, we propose a new complexity class for LQC, termed BLQP (bounded-error Lorentz quantum polynomial-time), encompassing problems solvable by LQC within polynomial time and with bounded errors. As the conventional quantum computer is
a special case of LQC, it is evident that BQP is a subset of BLQP.
We showcase LQC circuits capable of polynomially solving the NP-hard problem of finding the maximum independent set, thereby situating both NP and co-NP as subsets of BLQP. Our research further presents LQC algorithms that efficiently handle problems solvable in polynomial time within the complexity classes PP (probabilistic polynomial-time) and hence \pnp, establishing its equivalence to the \pnp complexity class.

A detailed comparison between LQC and quantum computing with postselection~\cite{Post} is presented, highlighting LQC's efficiency in simulating postselection and introducing a unique LQC capability termed super-postselection, which quantum computing with postselection cannot mimic. Consequently, the complexity class PostBQP, designated for quantum computing with postselection, is encompassed within BLQP.

We proceed with a concise review of LQC fundamentals and introduce two pivotal logic gates, the CV gate and CCV gate, essential for our effective algorithms tackling problems in NP, PP, and \pnp classes, and illustrating LQC's substantial edge over traditional quantum computing. The discussion concludes by contrasting LQC with quantum computing with postselection, further elucidating their relationship.

\section{Theoretical Model of Lorentz Quantum Computer}

In the referenced paper~\cite{LC}, a Lorentz quantum computer (LQC) model is detailed, drawing from the principles of Lorentz quantum mechanics~\cite{Pauli}, an extension of the Bogoliubov-de Gennes equation which describes bosonic Bogoliubov quasiparticle dynamics. A distinctive feature of these systems is their dual excitation branches, with only the bosonic Bogoliubov quasiparticles considered physically observable, while the negative energy counterpart is deemed unobservable~\cite{wunjp}. LQC capitalizes on this characteristic by introducing hyperbolic bits (or hybits for brevity), where one of its states is observable and the other is not. This concept aligns with prior studies involving systems with indefinite inner products~\cite{Bognar}, including work by Dirac and Pauli~\cite{Dirac,Pauli}.

In LQC, information storage involves two types of bits: conventional qubits and unique hybits. Qubits function
as they do in standard quantum computing, obeying  unitary transformations, while hybits are exclusive to LQC
and undergo complex Lorentz evolution under gate operations. The state of a hybit, denoted as $|\psi)$, is expressed as
\be
|\psi) = a|0) + b|1)=
\left(\begin{array}{c}
      a\\
      b\\
   \end{array}
   \right),
\ee
where $|0)$ and  $|1)$ are the computational bases satisfying
\be
   (0|\sigma_z|0) = 1 \,, ~~
   (1|\sigma_z|1) = -1 \,, ~~
   (1|\sigma_z|0) = 0 \,.
\ee
Here $\sigma_z = \text{diag}\{1,-1\}$ is the familiar Pauli $z$ matrix. In the following notation,
$|\thickspace)$ denotes the state of a hybit, while $|\thickspace\rangle$ denotes the state of conventional qubit.
Hybits $|\psi)$ evolve according to Lorentz quantum mechanics, maintaining a constant indefinite inner product over time
 \be
 \frac{d}{dt}(\psi|\sigma_z|\psi)=0\,.
 \ee
All the logic gates acting on a hybit induce Lorentz transformations, which preserve the indefinite inner product.
For example, if a hybit is in the state of  $|\psi)=(a,b)^T$,  after a gate operation $G$ it becomes $G|\psi)=(a',b')^T$, then we must have $|a'|^2-|b'|^2=|a|^2-|b|^2$. An important consequence is that there is no $\sigma_x$ operation that flips  between the hybit states $|0)$ and $|1)$,  because $(0|\sigma_z|0)=1$ and $(1|\sigma_z|1)=-1$.

Inherited from Lorentz quantum mechanics, for the two basis
of a hybit, only $|0)$ is observable  and $|1)$ is unobservable.
This is a fundamental and crucial property of the hybit;
as we will see later,  the power of LQC is largely derived from this feature. The extension to a multibit scenario
is straightforward (for a full elaboration, see Ref.~\cite{LC}).

Consider an LQC consisting of $N_q$ qubits and $N_h$ hybits.
Its state $|\Phi)$ can be expressed in the computational basis as
\be \label{Phi}
|\Phi)=\sum_{j=1}^{2^{N_q+N_h}}a_j|\psi_j)\,,
\ee
where
\ba
|\psi_j)&=&|q_1\rangle\otimes|q_2\rangle\cdots\otimes\ket{q_i}\cdots\otimes|q_{N_q}\rangle
\otimes|h_1)\otimes|h_2)\cdots\otimes|h_i)\cdots\otimes|h_{N_h})\nonumber\\
&=&|q_1,q_2\cdots q_i\cdots q_{N_q};h_1,h_2\cdots h_i\cdots h_{N_h})\,,
\ea
where $q_i$ and $h_i$ take values of either 0 or 1.

As long as $N_h\neq 0$, the LQC evolves according to the Lorentz transformation. Therefore, a multibit state must satisfy the indefinite inner product condition if at least one bit is a hybit. Consequently, a pure state of a system containing both qubits and hybits must be Lorentzian and is represented in the form $|\thickspace)$.

It is important to note that, if a term $|\psi_j)$  contains at least one $|1)$, it is not observable. For example, $|1,0\cdots,0;1,0,\cdots,0)$ is not observable. Also note that if $N_h = 0$ an LQC is reduced to a conventional quantum computer. In other words, a quantum computer is a special case of a Lorentz computer.

\begin{figure}[ht]
	\begin{subfigure}[b]{0.45\columnwidth}
		\centerline{~~~~\Qcircuit @C=1.2em @R=1.1em {
			\lstick{|\psi\rangle}  &   \gate{H}  & \qw  \\	
	 		\lstick{|\psi\rangle}   &   \gate{T}  & \qw
	 	}}
		\caption{}
  	\end{subfigure}
	\hspace{0.2cm}
	\begin{subfigure}[b]{0.45\columnwidth}
	\centerline{~~~~\Qcircuit @C=1.2em @R=1.1em {
			\lstick{|\psi)}   &   \gate{\tau}  & \qw  \\	
	 		\lstick{|\psi)}   & \gate{T}  & \qw
	 	}}
		\caption{}
	\end{subfigure}	
	\caption{(a) Single qubit gates $H$ and $T$; (b) single hybit gates $\tau$ and $T$.}
	\label{t2}
\end{figure}
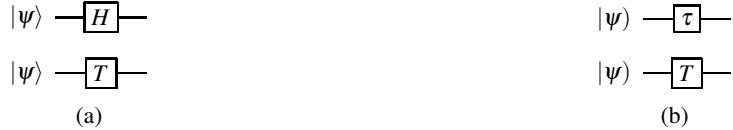

It has been established~\cite{LC} that the universal gates of an LQC consist of both single-bit gates
and two-bit gates in three distinct sets: $\{H, T\}$,
$\{\tau, T\}$, and $\{\Lambda_{1}^{qq}\left(\sigma_{z}\right), \Lambda_{1}^{qh}\left(\sigma_{z}\right),
\Lambda_{1}^{hq}\left(\sigma_{z}\right),\Lambda_{1}^{hh}\left(\sigma_{z}\right)\}$, where the subscript $1$ indicates that there is one control bit. The first set $\{H, T\}$ is the Hadamard gate $H$ and the $\pi/8$ gate $T$,
\be
	H = \frac{1}{\sqrt{2}}\left(\sigma_{x}+\sigma_{z}\right), \quad T = e^{-i\frac{\pi}{8}}
    \left(\begin{array}{cc}
	e^{i\pi/8} & 0 \\
	0 & e^{-i\pi/8}
	\end{array}\right) \,.
\ee
They are applicable to single qubits, and their combined application can approximate any single qubit
transformation with arbitrary precision. They are represented in circuits by the symbols in Fig.~\ref{t2}(a).

The second set operates on single hybits and consists of the $\pi/8$ gate $T$ and the  $\tau$ gate.
The $T$ gate has the same matrix form  as the $T$ gate for qubits, and the matrix form of the $\tau$ gate is
given by
\begin{eqnarray} \label{tau}
	\tau = \sqrt{2}\sigma_{z}+i\sigma_{x} = \left(\begin{array}{cc}
		\sqrt{2} & i \\
		i & -\sqrt{2}
		\end{array}\right) \,.
\end{eqnarray}
These two gates are applicable to single hybits. Their symbols in circuits are shown in Fig.~\ref{t2}(b).
It is noteworthy that the operator $H$ is unitary and $\tau$ is Lorentzian, while $T$ is both unitary and Lorentzian.
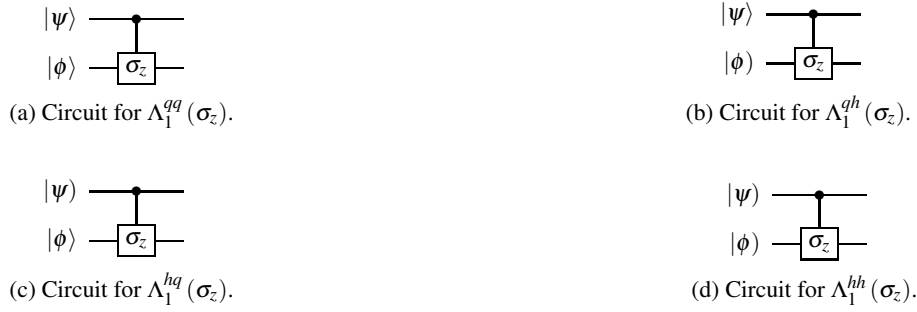
\begin{figure}[ht]

	\begin{subfigure}[b]{0.45\columnwidth}	
\centering
	~~~~~\Qcircuit @C=1.2em @R=1.3em {
				\lstick{|\psi\rangle}   &   \ctrl{1}     &   \qw      \\
				 \lstick{|\phi\rangle}   &   \gate{\sigma_z}   &    \qw       \\			 	
			 }
	\caption{Circuit for $\Lambda_{1}^{qq}\left(\sigma_{z}\right)$.}
	\end{subfigure}
	\hfill
	\begin{subfigure}[b]{0.45\columnwidth}
\centering
	~~~~~\Qcircuit @C=1.2em @R=1.3em {
				\lstick{|\psi\rangle}   &   \ctrl{1}     &   \qw      \\
				 \lstick{|\phi)}   &   \gate{\sigma_z}   &    \qw       \\			 	
			 }
	\caption{Circuit for $\Lambda_{1}^{qh}\left(\sigma_{z}\right)$.}
	\end{subfigure}
	\vspace{20pt}
	
	\begin{subfigure}[b]{0.45\columnwidth}
\centering
	~~~~~\Qcircuit @C=1.2em @R=1.3em {
				\lstick{|\psi)}   &   \ctrl{1}     &   \qw      \\
				 \lstick{|\phi\rangle}   &   \gate{\sigma_z}   &    \qw       \\			 	
			 }
	\caption{Circuit for $\Lambda_{1}^{hq}\left(\sigma_{z}\right)$.}
	\end{subfigure}
	\hfill
	\begin{subfigure}[b]{0.45\columnwidth}
\centering
	~~~~~\Qcircuit @C=1.2em @R=1.3em {
				\lstick{|\psi)}   &   \ctrl{1}     &   \qw      \\
				 \lstick{|\phi)}   &   \gate{\sigma_z}   &    \qw       \\			 	
			 }
	\caption{Circuit for $\Lambda_{1}^{hh}\left(\sigma_{z}\right)$.}
	\end{subfigure}
	\caption{Four different controlled-$\sigma_z$ gates}
	\label{t4}
\end{figure}

The logical gates in the final set, denoted as $\Lambda_{1}^{qq}\left(\sigma_{z}\right)$, $\Lambda_{1}^{qh}\left(\sigma_{z}\right)$, $\Lambda_{1}^{hq}\left(\sigma_{z}\right)$ and $\Lambda_{1}^{hh}\left(\sigma_{z}\right)$, represent four variations of controlled-$\sigma_{z}$ operators. These variations differ themselves by the types of the control and target bits as indicated by the superscripts: $q$ for qubit and $h$ for hybit. The corresponding circuits are illustrated in Fig.~\ref{t4}. Notably, we have chosen the controlled-$\sigma_z$ gate over the controlled-NOT (CNOT) gate, which is a more common choice in quantum computing. This decision is motivated by the fact that the CNOT gate is a unitary transformation, which does not hold for a hybit. In contrast, the controlled-$\sigma_z$ gate is both unitary and Lorentzian.

Note that the gates $\Lambda_{1}^{qq}\left(\sigma_{z}\right)$, $\Lambda_{1}^{qh}\left(\sigma_{z}\right)$, $\Lambda_{1}^{hq}\left(\sigma_{z}\right)$ and $\Lambda_{1}^{hh}\left(\sigma_{z}\right)$
are denoted in Ref.~\cite{LC} as $\Lambda_{1}^{qq}\left(\sigma_{z}\right)$, $\Lambda_{1}^{ql}\left(\sigma_{z}\right)$, $\Lambda_{1}^{lq}\left(\sigma_{z}\right)$ and $\Lambda_{1}^{ll}\left(\sigma_{z}\right)$, respectively.
The superscript $l$ is replaced by $h$ in this paper to avoid confusion.

It has been established~\cite{LC} that any Lorentz transformation of the state $|\Phi)$ in Eq.~(\ref{Phi}) can be realized by a combination of the gate sets $\{H, T\}$, $\{\tau, T\}$, and $\{\Lambda_{1}^{qq}\left(\sigma_{z}\right), \Lambda_{1}^{qh}\left(\sigma_{z}\right), \Lambda_{1}^{hq}\left(\sigma_{z}\right),\Lambda_{1}^{hh}\left(\sigma_{z}\right)\}$.

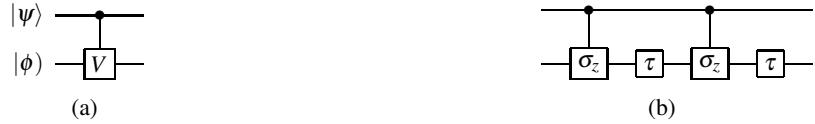
\begin{figure}[h!]	
\begin{subfigure}[b]{0.40\columnwidth}	
\centering
  		~~~~~\Qcircuit @C=1.2em @R=1.3em {
				\lstick{|\psi\rangle}   &   \ctrl{1}     &   \qw      \\
				 \lstick{|\phi)}   &   \gate{V}   &    \qw       \\			 	
			 }  	
	\caption{}
\end{subfigure}
\hspace{0.5cm}
\begin{subfigure}[b]{0.45\columnwidth}	
       \centering
  		~~~~~\Qcircuit @C=1.2em @R=1.5em {
				\lstick{}   &  \ctrl{1}     &    \qw       &      \ctrl{1} &           \qw &\qw    \\
				 \lstick{}   &   \gate{\sigma_z}&  \gate{\tau}   &   \gate{\sigma_z}       &  \gate{\tau}	&\qw		
}	
	\caption{}
\end{subfigure}
\caption{(a) Two-bit logical  $CV$ gate;
(b) a simple way to realize $CV$  using the controlled-$\sigma_z$ gate and the $\tau$ gate for
$\chi=2\ln(\sqrt{2}+1)$. }
	\label{t5}
\end{figure}

The following sections present powerful LQC algorithms for solving difficult problems. In these algorithms, one two-bit control gate is used repeatedly. It is the controlled-$V$ gate $\Lambda_{1}^{qh}\left(V\right)$, and we will denote it as CV. Its circuit is shown in Fig.~\ref{t5}(a), where the control bit is a qubit and the target bit is a hybit. If the qubit is in the state of $\ket{0}$, nothing happens; if it is in the state of $\ket{1}$, the hybit undergoes a complex Lorentz transformation
\begin{equation} \label{Ltransformation}
V=\left(\begin{array}{cc}\cosh \chi&-\text{i}\sinh\chi\\ \text{i}\sinh\chi& \cosh\chi  \end{array}\right),
\end{equation}
where $\chi$ is a positive constant. The transformation $V$ is actually a hyperbolic rotation: for a positive integer $r$, we have
\begin{equation} \label{Ltransformation2}
V^r=\left(\begin{array}{cc}\cosh r\chi&-\text{i}\sinh r\chi\\ \text{i}\sinh r\chi& \cosh r\chi  \end{array}\right)\,.
\end{equation}
For $\chi=2\ln(\sqrt{2}+1)$, as shown in Fig.~\ref{t5}(b),
the CV gate can be realized with two $\tau$ gates and two controlled-$\sigma_z$ gates.

\begin{figure}[h!]

\flushleft{(a)}\\ \centering
		\hspace{0cm}\Qcircuit @C=1.2em @R=1.3em {
				\lstick{|\psi_1\rangle}   &   \ctrl{1}     &   \qw      \\
                \lstick{|\psi_2\rangle}   &   \ctrl{1}     &   \qw      \\
				 \lstick{|\phi)}   &   \gate{V}   &    \qw       \\			 	
			 } 	
\vspace{0.2cm}
\flushleft{(b)}\\ \centering
  		\hspace{0mm}\Qcircuit @C=1.2em @R=1.3em {
\lstick{|\psi_1\rangle}& \ctrl{2}  & \qw   &\qw           &   \qw     &\ctrl{2}        &\qw        & \qw &\qw  &\qw \\
\lstick{|\psi_2\rangle}& \qw      &  \qw   &\ctrl{1}      &   \qw     &\qw             &\qw &  \ctrl{1} &\qw &\qw\\
\lstick{|\phi)}&\gate{\sigma_z}&\gate{\tau}&\gate{\sigma_z}&\gate{\tau}&\gate{\sigma_z}&\gate{\tau}&\gate{\sigma_z} &\gate{\tau}    &\qw        \\				
}	

\caption{(a) Three-bit logical  CCV gate; (b) the circuit that implements the CCV gate with four $\tau$ gates and
	four controlled-$\sigma_z$ gates for $\chi=4\ln(\sqrt{2}+1)$. }
\label{t51}
\end{figure}
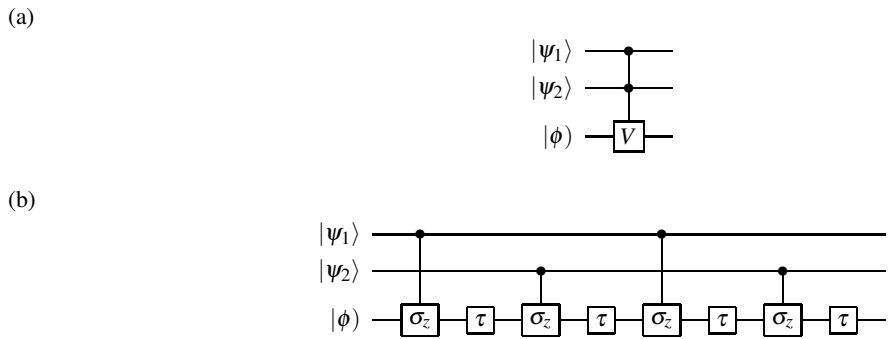

We also often use a three-bit logic gate as shown in Fig.~\ref{t51}(a),
where the two qubits are control bits and the hybit is the target bit. Only when both the qubits are in the state $|1\rangle$, does the target hybit undergo the Lorentz transformation $V$; otherwise nothing happens.
We call this a CCV gate, which can be realized with a circuit in Fig.~\ref{t51}(b). This circuit consists of four $\tau$ gates and four controlled-$\sigma_z$ gates for $\chi=4\ln(\sqrt{2}+1)$.

The  CV and CCV gates are at the heart of LQC's computational power since the Lorentz transformation in Eq.~(\ref{Ltransformation}) has the ability to amplify the components of a hybit state without limit. Consider a system of a qubit and a hybit that is in the state of
\be
|\phi_0)=\frac{\sqrt{2}}{2}\big[\ket{0}+\ket{1}\big]\otimes |0)\,.
\ee
After the application of a CV gate, the state becomes
\be
\frac{\sqrt{2}}{2}\ket{0}\otimes |0)+\frac{\sqrt{2}}{2}\ket{1}\otimes \big[\cosh\chi |0)+\text{i}\sinh\chi |1)\big]
\ee
As mentioned before, the state $|1)$ for a hybit is unobservable so that we only need to
consider the two terms that contain $|0)$, which are
\be
|\phi_1)=\frac{\sqrt{2}}{2}\big[\ket{0}+\cosh\chi\ket{1}\big]\otimes  |0)\,.
\ee
Compared to the state $|\phi_0)$, it is clear that the weight of the state $\ket{1}$ has increased in both absolute and relative terms. As we will see in the following sections, this capability of the gate CV gives LQC a significant computational advantage over the conventional quantum computer. The gate CCV has a similar ability to selectively amplify. With this capacity of amplification in mind, we introduce the formal definition of BLQP, a computational complexity class of languages related to LQC.

\textbf{\it Definition of BLQP}. For a language L within BLQP, there exists a uniform family of quantum circuits, denoted as $\{\mathbb{C}_n\}_{n\geq1}$, where each circuit is of polynomial size. These circuits employ qubits and hybits, as well as unitary and Lorentzian gates, and they allow measurements after which no further quantum gates can be applied. Given an input of length $n$ and specific initial states for the work qubits and hybits, the circuit $\mathbb{C}_n$ operates for polynomial time in $n$ and then halts. For $\omega\in\text L$, the probability of obtaining an accepting state is greater than $2/3$. Conversely, for $\omega\notin\text L$, this probability is less than $1/3$.

The criterion for an accepting state can involve either all bits in $\mathbb{C}_n$ or a single qubit. For instance, a specific qubit, referred to as the ``Y qubit," can be used for this purpose. An accepting state is defined as:
\be
|\text{accept})=|\Psi_1\rangle\otimes|00\ldots0)\otimes|1_Y\rangle,
\ee
while a rejecting state is:
\be
|\text{reject})=|\Psi_2\rangle\otimes|00\ldots0)\otimes|0_Y\rangle,
\ee
where $|\Psi_1\rangle$ and $|\Psi_2\rangle$ (with $\langle\Psi_1|\Psi_1\rangle=\langle\Psi_2|\Psi_2\rangle=1$) represent the states of all qubits determined by the circuit's output. The term $|00\ldots0)$ indicates that all hybits are in the state $|0)$, where $|0)$ is detectable and $|1)$ is undetectable. The subscript $Y$ in $|1_Y\rangle$ and $|0_Y\rangle$ denotes the state of the ``Y qubit".

According to the properties of hybits, for $\omega\in\text L$, the output of $\mathbb{C}_n$ will be of the form:
\be \label{ZZZZ}
|\psi)=c_{\text{yes}}|\Psi_1\rangle\otimes|00\ldots0)\otimes|1_Y\rangle+c_{\text{no}}|\Psi_2\rangle\otimes|00\ldots0)\otimes|0_Y\rangle+c_3|\Psi_3),
\ee
where $\frac{c_{\text{yes}}^2}{c_{\text{yes}}^2+c_{\text{no}}^2}>2/3$. Here, $|\Psi_3)$ represents the overall state associated with all bits in the circuit when at least one hybit is in the undetectable state $|1)$.

Similarly, for an input $\omega\notin\text L$ of length $n$, the output of $\mathbb{C}_n$ will also be in the form of the expression above, but with $\frac{c_{\text{yes}}^2}{c_{\text{yes}}^2+c_{\text{no}}^2}<1/3$.

For convenience, this error probability is often expressed as an exponentially small quantity rather than using $1/3$.

\section{LQC algorithms for the maximum independent set}
\label{sec:mis}
In this section, we will introduce an LQC algorithm capable of solving the maximum independent set (MIS) problem in polynomial time. Given that MIS is NP-hard~\cite{Xiao}, this directly implies that LQC can polynomially solve all problems within the NP and co-NP classes. It is noteworthy that MIS does not belong to PP, indicating that BLQP is not encompassed by PP.

For a graph $G(n,m)$ with $n$ vertices and $m$ edges, an independent set (IS) is a subset of the vertices that are not directly connected by edges. The maximum independent sets (MIS) are those with the largest number of vertices among
all ISs. For a given graph, finding its MIS is difficult on a classical computer and it is a NP-hard problem~\cite{Xiao}. Moreover, for a given graph $G(n,m)$, no classical algorithm can find an appropriately approximate
MIS in polynomial time in the worst case~\cite{Johan,Zuckerman}. For many graphs, the largest IS found by polynomial-time classical algorithms is only about half the size of the MIS~\cite{Coja,Frieze}. A recently proposed quantum algorithm shows promising signs of exponential speedup~\cite{Yu2021,WYW}; however, there is no rigorous proof or very convincing numerical evidence. Here we present an LQC algorithm that can solve MIS problems in polynomial time.

To design the algorithm for a given graph $G(n,m)$, we assign a Boolean variable to each vertex, $x_1,x_2,\cdots,x_n$.  As a result, a subset of the vertices is represented by an integer $x$ in its $n$-digit binary form: if its $i$th digit $x_i=1$ then the $i$th vertex is in the subset; $x_i=0$ then it is not. If $x$ is an IS, then its $x_i$ and $x_j$ cannot both be $1$ simultaneously if the two vertices $x_i$ and $x_j$ are connected by an edge.

For an LQC algorithm, we use $n$ work qubits to represent the $n$ vertices. Their $N=2^n$ possible states $|00...0\rangle$, $|00...1\rangle$, ..., $|11...1\rangle$ naturally represent all the subsets of vertices. That is, a basis vector $\ket{x}$ corresponds to the subset $x$ where the integer $x$ is in its binary form.
The goal is to find the target state $|M\rangle$ that corresponds to MIS out of the $N=2^n$ possible states.

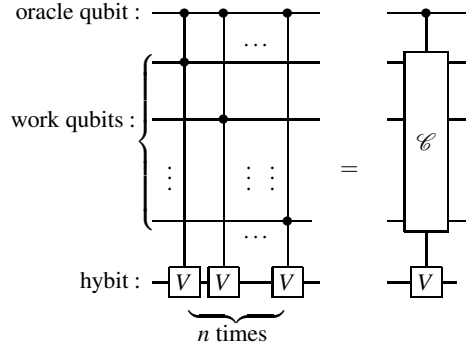
\begin{figure}[h!]
\hspace*{0.5cm}
	\centerline{
  \Qcircuit @C=0.35em @R=1.5em {
   \lstick{\rm oracle\ qubit : } &\qw & \ctrl{1}& \ctrl{2}& \qw &\qw &\qw& \ctrl{4} & \qw&\qw& & && \qw&\ctrl{1} &\qw&\qw\\
  \lstick{} &\qw & \ctrl{4}& \qw & \qw & \ustick{\cdots }\qw & \qw  & \qw & \qw & \qw & & &  & \qw&\multigate{3}{{\mathcal C}}&\qw&\qw\\
  \lstick{} &\qw & \qw & \ctrl{3} & \qw & \qw & \qw & \qw &  \qw & \qw & & & &\qw& \ghost{{\mathcal C}}&\qw&\qw\\
  \lstick{} &    & \lstick{\vdots} & &   & & \lstick{\vdots } & \lstick{\vdots } & &  &\push{\rule{.3em}{0em}=\rule{.3em}{0em}}  &  & & & &&\\
  \lstick{} &\qw &\qw & \qw &\qw & \dstick{\cdots } \qw&\qw&  \ctrl{1}& \qw  &&  & & &\qw&\ghost{{\mathcal C}}&\qw&\qw
	 \inputgroupv{2}{5}{0.6em}{2.5em}{{\rm work~qubits:} ~~~~~~~~~~~~~}\\
	 \lstick{{\rm hybit} :}  & \qw & \gate{V} &\gate{V} & \qw  &\qw     &\qw & \gate{V}  & \qw  &  \qw &&&  &\qw&\gate{V}\qwx&\qw\qw
	 \gategroup{6}{4}{6}{7}{1.6em}{_\}}\\
	 & & &  \mbox{~~~$n$ times} &&&& &  & &&&&&&
  }
  }
	\caption{LQC circuit for the operation $Q$ that is capable of counting
	the number of ones that are in the basis state $\ket{x}$. It
	consists of $n$ CCV gates.
	The right circuit is a short-hand representation of the left circuit.}
	\label{QC}
\end{figure}

In our algorithm for MIS problems, we add an oracle qubit and a hybit on top of the $n$ work qubit in the computation circuit. The main part of our algorithm is shown in Fig.~\ref{QC}, which consists of $n$ CCV gates. To see its functionality, let us consider two basis states $\ket{x}$ and $\ket{y}$: $x$ is not an IS and $y$ is an IS.
To distinguish them, we entangle them with the oracle qubit and prepare the following initial state
\be
\label{phi0}
|\phi_0)=\frac{1}{\sqrt{2}}(\ket{x}\otimes\ket{0_o}+\ket{y}\otimes\ket{1_o})\otimes |0)\,.
\ee
The $Q$ operation shown in Fig.~\ref{QC} consists of $n$ CCV gates. After its application, the state at the output is
\begin{align}
|\phi_1)=&Q|\phi_0)=
\frac{1}{\sqrt{2}}\Big[\ket{x}\otimes\ket{0_o}\otimes |0)
+\cosh(m_y\chi)\ket{y}\otimes\ket{1_o}\otimes |0)
+\text{i}\sinh(m_y\chi)\ket{y}\otimes\ket{1_o}\otimes |1)\Big]\,,
\end{align}
where $m_y$ is the number of ones in the binary form of $y$. As emphasized in the last section, the hybit state $|1)$ is unobservable. So, the above state effectively has only the first two terms. As a result, the weight of the state $\ket{y}$ is enhanced by a factor of $\cosh(m_y\chi)$, which is determined by $m_y$, the number of ones in $y$. This means that the circuit in Fig.~\ref{QC} effectively has the ability to count the number of ones in $y$, which for the graph is the number of vertices in the subset $y$.

To achieve an entangled state similar to $|\phi_0)$ in Eq.(\ref{phi0}), we use the following oracle
\begin{equation} \label{misOracle}
O_{\rm IS}=(I-P_{\rm IS})\otimes I_o+P_{\rm IS}\otimes (|0_o\rangle\langle1_o|+|1_o\rangle\langle 0_o|)\,,
\end{equation}
where $I_o$ is the identity matrix for the oracle qubit and $P_{\rm IS}$ is a projection onto the sub-Hilbert space spanned by all possible solutions $|j\rangle$ of IS,
\be
P_{\rm IS}=\sum_{x\in {\rm IS}}|x\rangle\langle x|\,.
\ee
The quantum oracle $O_{\rm IS}$ is similar to the one used in the Grover algorithm~\cite{Nielson}, and it evaluates whether a subset $x$ is an IS in polynomial time.

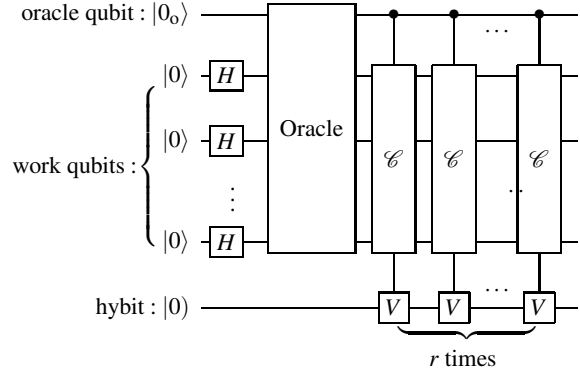
\begin{figure}[t]
\hspace*{0.5cm}
	\centerline{
  \Qcircuit @C=0.35em @R=1.5em {
  \lstick{\rm oracle\ qubit : \ket{0_o}}  & \qw &\qw & \qw &   \multigate{4}{{\rm Oracle}}  &  \qw
	 &\ctrl{1}&\qw&\ctrl{1}&\qw&\dstick{~\cdots}\qw&\qw&\qw&\ctrl{1}&\qw&\qw\\
	 \lstick{\ket{0}} & \gate{H} & \qw& \qw  &   \ghost{{\rm Oracle}}  & \qw&\multigate{3}{{\mathcal C}}&\qw&\multigate{3}{{\mathcal C}}&\qw&\qw&\qw&\qw&\multigate{3}{{\mathcal C}}&\qw&\qw\\
	 \lstick{\ket{0}}  &  \gate{H} & \qw & \qw &   \ghost{{\rm Oracle}}  & \qw &  \ghost{{\mathcal C}}&\qw&\ghost{{\mathcal C}}&\qw&\qw&\qw&\qw&\ghost{{\mathcal C}}&\qw&\qw\\
	 \lstick{}   &    &  &  \lstick{\vdots~~}   &    &   & &&&&&&&&\lstick{\cdot\cdot~~~~~}&\\     
	 \lstick{\ket{0}}  &   \gate{H}   & \qw &\qw & \ghost{{\rm Oracle}}  & \qw  &  \ghost{{\mathcal C}}&\qw
	 &\ghost{{\mathcal C}}&\qw  &\qw &\qw&\qw
	 &\ghost{{\mathcal C}}&\qw &\qw
	 \inputgroupv{3}{4}{4.5em}{1.1em}{{\rm work~qubits:} ~~~~~~~~~~~~~~~~~~~~~~~~~~~~~}\\
	 \lstick{{\rm hybit} : |0)}  & \qw  & \qw &\qw   &   \qw  &  \qw &\gate{V}\qwx&\qw&\gate{V}\qwx &\qw
	 &\ustick{~\cdots}\qw &\qw&\qw&\gate{V}\qwx &\qw&\qw
	 \gategroup{6}{8}{6}{13}{2.1em}{_\}}\\
	 & & &  & &  &&&\mbox{~~~$r$ times} &&&& &&&&
	   }
  }
\caption{Circuit of an LQC algorithm for solving MIS problems in polynomial time. The big box represents the oracle that implements the operator ($\ref{misOracle}$). As explained in the text, $r$ is proportional to $n$.  }
\label{t7}
\end{figure}

The circuit of our algorithm is shown in Fig.~\ref{t7}.
The initial state of the whole system, including the $n$ work qubits, one oracle qubit and one hybit, is set to be $|00\ldots0\rangle\otimes|0_o\rangle\otimes|0)$. The algorithm then proceeds as follows:

(i) apply Hadamard gates to all work qubits;

(ii) apply the oracle $O_{\rm IS}$;

(iii) apply the $Q$ operation $r$ times;

(iv) measure the oracle qubit and the hybit.

After the step (i), the state becomes
\begin{align}
\label{iniH}
|\Psi_0)=|\Phi_0\rangle\otimes|0_o\rangle
\otimes |0)
=\frac{1}{\sqrt{N}}\left(\sum_{x=0}^{2^n-1}|x\rangle\right)\otimes|0_o\rangle\otimes|0)\,.
\end{align}
With the oracle operation in the step (ii), we have
\begin{align}
|\Psi_1)=&O_{\rm IS}|\Phi_0\rangle\otimes
|0_o\rangle\otimes|0)
=\left(\sum_{j\notin\text{IS}}|j\rangle\otimes|0_o\rangle
+\sum_{x\in\text{IS}}|x\rangle\otimes|1_o\rangle \right)\otimes|0)\,.
\end{align}
 After the step (iii), we obtain
\begin{align}
|\Psi_2)=Q^r|\Psi_1) &=\Big(\sum_{j\notin\text{IS}}\ket{j}\otimes\ket{0_o}
+\sum_{x\in\text{IS}}\cosh(m_xr\chi)\ket{x}\otimes\ket{1_o}\Big)\otimes|0)\nonumber\\
&+ \Big(\sum_{x\in\text{IS}}\text{i}\sinh(m_x r\chi)\ket{x}\otimes\ket{1_o}\Big)\otimes|1),
\end{align}
where $m_x$ is the number of ones in the binary form of $x$, or equivalently, the number of vertices in the IS $x$.
As  mentioned in the last section, the hybit state $|1)$ is not observable. So, the probability $P$ of getting the MIS after the measurement is given by
\begin{equation}
\label{misp}
P=\frac{N_{\rm MIS}\cosh^2(Mr\chi)}{N-N_{\text{IS}}+\sum_{x\in\text{IS}}\cosh^2(m_xr\chi)},
\end{equation}
where $M$ is the number of vertices in the MIS, $N_{\text{IS}}$ is the number of ISs,
and $N_{\text{MIS}}$ is the number of  MIS. It is obvious that we have
\begin{eqnarray}
P>\frac{N_{\text{MIS}}\cosh^2(Mr\chi)}{(N-N_{\text{MIS}})\cosh^2((M-1)r\chi)+N_{\text{MIS}}\cosh^2(Mr\chi)}
\approx\frac{N_{\text{MIS}}e^{2r\chi}}{N-N_{\text{MIS}}+N_{\text{MIS}}e^{2r\chi}}.
\end{eqnarray}
It is clear that $P\approx1$ when $r\approx\frac{1}{\chi}\ln{N}\propto n$. Since each execution of $Q$ involves $n$
CCV gates, the time complexity of our algorithm is $O(n\ln{N})\sim O(n^2)$.

The definition of BLQP is inherently linked to decision problems. By trivially extending the circuit, we can transform the problem into a decision problem. The input for the decision algorithm is of the form ``$G(n,m)$ (a graph with $n$ vertices and $m$ edges) + $S$ (a subset of vertices in $G$)", and the output is of the form shown in Eq.~(\ref{ZZZZ}),
\be
|\psi)=c_{\text{yes}}|\text{yes})+c_{\text{no}}|\text{no})+c_3|\Psi_3),
\ee
where $\frac{c_{\text{yes}}^2}{c_{\text{yes}}^2+c_{\text{no}}^2}>2/3$ if $S$ forms an MIS in $G$, and $\frac{c_{\text{yes}}^2}{c_{\text{yes}}^2+c_{\text{no}}^2}<1/3$ if $S$  does not form an MIS. $|\Psi_3)$ represents the overall state associated with all bits in the circuit when at least one hybit is in the undetectable state $|1)$.

This extension to the decision circuit can be easily achieved by adding an additional oracle to the original circuit, as shown by the small box in Fig.~7. The portion of input $G(n,m)$ is used for the initial circuit in Fig.~7, while the portion of $S$ serves as the input for the additional oracle. By checking whether the output of the circuit in Fig.~7 matches $S$ by the additional oracle, we can solve this decision problem with the oracle qubit for the additional oracle functioning as the ``Y qubit", as indicated in Eq.~(\ref{ZZZZ}).


It is clear that by extending the circuit above for this decision problem, $|c_{\text{yes}}|^2$​​ will be nearly 1 if $S$ is an MIS, and nearly 0 if $S$ is not. Since LQC can solve the MIS problem, which is NP-hard, in polynomial time, it follows that both NP and co-NP are subsets of BLQP.

In fact, the MIS problem is also in the ${\text P}^{\text{NP}}$ complexity class belonging to PH (polynomial hierarchy). The $k$-IS, which involves finding an independent set of $k$ vertices, falls within NP. We can submit $n$ non-adaptive queries to the NP-oracle (or SAT-oracle since the SAT problem is NP-complete) for $0$-IS, $1$-IS, all the way up to $n$-IS solutions. By determining the maximum value of $k_{\text{MAX}}$ that yields a positive result from the oracle, we derive the solution for MIS. Thus, MIS is in the class ${\text P}^{\parallel\text{NP}}$​​~\cite{Buss,Hemachandra}.

\section{\pnp$\subseteq$BLQP}
In this section, we will first present an LQC algorithm that solves problems of the complexity class PP in polynomial time. We will then show that this algorithm can be used to solve problems in the class of ${\text P}^{\sharp \text{P}}$ in polynomial time. This means that both PP and ${\text P}^{\sharp \text{P}}$ are subsets of BLQP.

\subsection{LQC algorithm for PP}

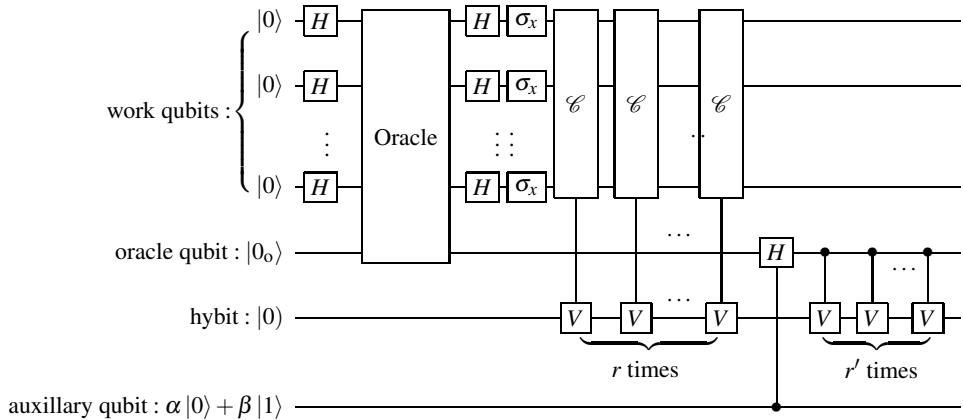
\begin{figure*}[th!]
\hspace*{0.5cm}
	\centerline{
  \Qcircuit @C=0.35em @R=1.5em {
	 \lstick{\ket{0}} & \gate{H} & \qw& \qw  &   \multigate{4}{{\rm Oracle}}  & \qw&\gate{H}&\gate{\sigma_x}&\multigate{3}{{\mathcal C}}&\qw&\multigate{3}{{\mathcal C}}&\qw&\qw&\qw&\qw&\multigate{3}{{\mathcal C}}&\qw&\qw&\qw&\qw&\qw&\qw&\qw&\qw&\qw&\qw&\qw\\
	 \lstick{\ket{0}}  &  \gate{H} & \qw & \qw &   \ghost{{\rm Oracle}}  & \qw &\gate{H}&\gate{\sigma_x}&  \ghost{{\mathcal C}}&\qw&\ghost{{\mathcal C}}&\qw&\qw&\qw&\qw&\ghost{{\mathcal C}}&\qw&\qw&\qw&\qw&\qw&\qw&\qw&\qw&\qw&\qw&\qw\\
	 \lstick{}   &    &  &  \lstick{\vdots~~}   &    &&\rstick{\vdots~~~}& \lstick{~~~\vdots} & &&&&&&&&\lstick{\cdot\cdot~~~~~}&&&&&&&&&&\\     
	 \lstick{\ket{0}}  &   \gate{H}   & \qw &\qw & \ghost{{\rm Oracle}}  & \qw  &\gate{H}&\gate{\sigma_x}&  \ghost{{\mathcal C}}&\qw
	 &\ghost{{\mathcal C}}&\qw  &\qw &\qw&\qw
	 &\ghost{{\mathcal C}}&\qw &\qw&\qw&\qw&\qw&\qw&\qw&\qw&\qw&\qw&\qw
	 \inputgroupv{2}{3}{4.5em}{1.1em}{{\rm work~qubits:}  ~~~~~~~~~~~~~~~~~~~~~~~~~~~~~}\\
	 \lstick{\rm oracle\ qubit : \ket{0_o}}  & \qw &\qw & \qw &   \ghost{{\rm Oracle}}  &  \qw  &\qw&\qw
	 &\qw\qwx&\qw&\qw\qwx&\qw&\ustick{~\cdots}\qw&\qw&\qw&\qw\qwx&\qw&\gate{H}&\qw&\ctrl{1}&\qw&\ctrl{1}&\qw&\dstick{\cdots}\qw&\ctrl{1}&\qw&\qw\\
	 \lstick{{\rm hybit} : |0)}  & \qw  & \qw &\qw & \qw & \qw &\qw&\qw &\gate{V}\qwx&\qw&\gate{V}\qwx &\qw
	 &\ustick{~\cdots}\qw &\qw&\qw&\gate{V}\qwx &\qw&\qw&\qw&\gate{V}&\qw&\gate{V}&\qw&\qw&\gate{V}&\qw&\qw
	 \gategroup{6}{10}{6}{15}{2.1em}{_\}}\gategroup{6}{21}{6}{24}{2.1em}{_\}}\\
	 & & &  & &  &&&&&\mbox{~~~$r$ times} &&&& &&&&&&&\mbox{~~~$r'$ times}&&&\\
	 \lstick{\rm auxillary\ qubit : \alpha\ket{0}+\beta\ket{1}}  & \qw &\qw & \qw & \qw & \qw &\qw&\qw
	 &\qw &\qw&\qw&\qw&\qw&\qw&\qw&\qw&\qw&\ctrl{-3}&\qw&\qw&\qw&\qw&\qw&\qw&\qw&\qw&\qw
	   }
  }
     \hspace*{2.5cm}\vspace{20pt}
	\caption{Circuit of an LQC algorithm for solving MAJSAT, which is PP-complete. The auxiliary qubit is initialized in $\alpha|0\rangle+\beta|1\rangle$ with $\beta/\alpha=2^i$, where $i$ is an integer ranging from $-n$ to $n$.}
\label{t9}
\end{figure*}

Instead of considering the class of PP (probabilistic polynomial-time) problems in general, we focus on a PP-complete problem, MAJSAT, and discuss the approach to solving it with LQC. For a given Boolean expression $f(x_1,x_2,\cdots,x_n)$ of $n$ Boolean variables, the problem of MAJSAT is to determine whether major assignments of Boolean variables satisfy $f=1$. To understand why MAJSAT is PP-complete, please consult some textbooks on computational complexity theory, for example, Ref.~\cite{Arora}. For a given Boolean formula $f$, we let $s$ be the number of assignments of $n$ Boolean variables satisfying $f=1$. The problem of MAJSAT is to determine whether $s> 2^{n-1}$.

The LQC circuit for solving MAJSAT is depicted in Fig.~\ref{t9}, where an auxiliary qubit is used together
with $n$ work qubit, an oracle qubit and a hybit. The quantum oracle used here is similar to the one in Fig.~\ref{t7} and has the ability to evaluate $f$ in parallel in polynomial time . We continue to use binary notation, i.e., in a state $\ket{x}$ of
$n$ work qubit, the integer $x$ is understood in its binary form.  The algorithmic steps as shown in  Fig.~\ref{t9} are as follows:

(i) Initialize all  bits  to either $\ket{0}$ or $|0)$ except the auxiliary qubit, which
is set to the state of  $\ket{\varphi_{\beta/\alpha}}=\alpha|0\rangle+\beta|1\rangle$
 where $\alpha$ and $\beta$ are real and positive.

(ii) Apply Hadamard gates to each of the work qubits,
\be
|\Psi_{\rm ii})=|\Phi_0\rangle\otimes|0_o\rangle\otimes|0)\otimes\ket{\varphi_{\beta/\alpha}}
=\frac{1}{\sqrt{N}}\left(\sum_{j=0}^{2^{n}-1}|j\rangle\right)\otimes|0_o\rangle\otimes|0)\otimes\ket{\varphi_{\beta/\alpha}}.
\ee

(iii) Apply the oracle operator $O$  to the state vector,
\begin{equation} \label{Oracle}
O=(I-P_s)\otimes I_o+P_s\otimes (|0_o\rangle\langle1_o|+|1_o\rangle\langle 0_o|)\,,
\end{equation}
where $I_o$ is the identity matrix for the oracle qubit and $P_s$ is a projection onto the sub-Hilbert space spanned by all possible solutions $|j\rangle$ of $f=1$
\be
P_s=\sum_{j\in \{f=1\}  }|j\rangle\langle j|\,.
\ee
followed by applying Hadamard gates and $\sigma_x$-gates to each of the work qubits.

(iv) Apply the  $Q$ operation (see Fig.~\ref{QC}) $r$ times without using the oracle qubit as the
control bit.

(v) Apply the Hadamard gate to the oracle qubit with the auxiliary qubit as the control bit.

(vi) With the oracle qubit as the control bit and the hybit as the target bit, apply
the CV gates $r'$ times.

 (vii) Measure the auxiliary qubit in the $x$-direction a large number of times, and
 count the number of one of the two outcomes: $1$ and $-1$. Note that
 one needs to repeat steps (i)-(vi) for each measurement. The number of
 measurement  will be discussed in later analysis.

(viii) Repeat the above procedures $2n+1$ times, each with a distinct value of $\beta/\alpha=2^i$
for the auxiliary qubit, where $i$ is an integer ranging from $-n$ to $n$ inclusive.

Let us analyze the algorithm to understand why it is capable of solving MAJSAT. After step (iii),
the entire system becomes
\be
|\Psi_{\rm iii})=\sum_{x=0}^{N-1} \ket{x}\otimes
\big(a_x\ket{0_o}+b_x\ket{1_o}\big)\otimes |0)
\otimes \ket{\varphi_{\beta/\alpha}}\, ,
\ee
where $a_x=1$ and $b_x=0$ when $f(x_1,x_2,\cdots,x_n)=0$, and $a_x=0$ and $b_x=1$ when $f(x_1,x_2,\cdots,x_n)=1$. With $r\approx \ln N/\chi$, the subsequent $Q$ operations in step (iv) are aimed at effectively isolating the term with $\ket{11\cdots 1}$ among all the possible $N=2^n$ terms.
Omitting terms with exponentially smaller coefficients and terms with $|1)$ which are unobservable,
we have
  \be
|\Psi_{\rm iv})\approx \ket{11\cdots 1}\otimes
\ket{\phi_o}\otimes |0)
\otimes \ket{\varphi_{\beta/\alpha}}\,,
\ee
where
\be
\ket{\phi_o}=\frac{(N-s)\ket{0_o}+s\ket{1_o}}{\sqrt{(N-s)^2+s^2}}\,.
\ee
In this step, the number of satisfying assignments $s$ is stored in the coefficients of
the oracle qubit state $\ket{\phi_o}$. Since the difference between the two coefficients can be
exponentially small,  in general one has  to measure it exponentially many times
in order to tell the difference. Steps (v) and (vi)  use the special property of hybit to reduce
it to polynomial number of times.

At the step (v), the controlled Hadamard gate mixes up the coefficients of the oracle qubit state and the auxiliary qubit state.
At the step (vi),  the CV gate is applied $r'$ times  with the oracle qubit as the control bit and the hybit as the target bit.
By setting $r'\approx \ln N/\chi$, when we measure the oracle qubit, we are
almost certain to find it in the state of $\ket{1_o}$ and the auxiliary qubit in the state of
\begin{equation} \label{PPS}
|\varphi_{\eta}\rangle=\frac{ s|0\rangle+\eta\sqrt{1/2}(2^n-2s)|1\rangle}{\sqrt{s^2+(\eta^2/2)(2^n-2s)^2}}\,,
\end{equation}
where $\eta=\beta/\alpha$. The detailed calculation leading to the above equation can be found in Appendix B.
As indicated in the step (viii),
$\eta$ has $2n+1$ possible values, $\eta_i=2^i$ ($i\in [-n,n]$).

With $|\varphi_{\eta}\rangle$, it is now possible to determine
whether $s>2^{n-1}$  in polynomial time.  We regard the auxiliary qubit as a spin, and  measure
it along the $x$-direction for which the two basis vectors are $|\pm\rangle=(|0\rangle\pm|1\rangle)/\sqrt{2}$.
When $s\leq 2^{n-1}$, it is easy to show that
\begin{equation} \label{ProPP2}
P_-=|\langle-|\varphi_{2^i}\rangle|\leq1/2.
\end{equation}
This means that if we measure it large number of times,
the number of outcome $-1$ will not exceed the number of outcome $1$.

When the instance $f$ is in MAJSAT, that is,  $s>2^{n-1}$, the
probability of outcome  $-1$ is
\be
P_-=\frac12+\frac{\sqrt{2}\eta s(2s-2^n)}{2s^2+\eta^2(2s-2^n)^2}\,,
\ee
which is always greater than $1/2$.  Although $\delta_p=P_--1/2$
can be exponentially small for some values of $\eta$,
we find that for a given value of $\delta_p\leq \sqrt{2}/4$,
there are some values of $\eta$ such that for all  possible
values of $s>2^{n-1}$  we always have
\begin{equation} \label{ProPP}
P_-\geq\frac{1}{2}+\delta_p\,.
\end{equation}
For convenience, we denote one of such $\eta$ as $\eta_a=2^{m_a}$,
where $m_a$ is an integer between $-n$ and $n$.
For detailed analysis, please see Appendix C.
Note that  $\delta_p$ can be set to a small value but the value is  finite and  independent of $n$.
For a smaller $\delta_p$, there are more possible values of $\eta$, similar to $\eta_a$.
Suppose for the special $\eta_a$, the auxiliary qubit is measured $N_{\text{PP}}$~times.
We find that when
\begin{equation}
\label{PPnumber}
N_{\text{PP}}\geq \frac{2\log(\epsilon)}{\log(1-4\delta_p^2)}\,,
\end{equation}
the probability that the number of measurement results $-1$  exceeds the number of results $1$
for  $\eta_a$ is $P=1-\epsilon$.

Based on the above analysis, we lay out the procedure
to determine if $f$ is in MAJSAT. First, we  set $\epsilon=c^{-n}$ with $c>1$. This makes the value
of $N_{\text{PP}}$ becomes a linear function of $n$.
Then, for each value of $\eta_i=2^i$ ($i\in[-n,n]$), we conduct $n$ sets of measurements. Within each set, we perform $N_{\text{PP}}$ measurements. This means that, for any given value of $\eta_i=2^i$, the auxiliary qubit is measured in the $x$-direction a total of $nN_{\text{PP}}$ times.
For each set of $N_{\text{PP}}$ measurements,
designate a result as ``success" if the number of results $-1$ exceeds the number of results $1$.
For the $n$ sets of $N_{\text{PP}}$ measurements, count the occurrences of ``success" results.

If $f$ is in MAJSAT, then the probability of all the results being ``success" at $\eta_a$ is given by
\begin{equation}
P_n=(1-c^{-n})^n\,.
\end{equation}
It is evident that $\lim_{n\rightarrow\infty}P_n=1$. This means that, for a sufficiently large $n$, there must exist a value of $\eta=\beta/\alpha=2^i$ such that the results of all $n$ sets of measures are ``success".

However, if $f$ is not in MAJSAT, then the probability of having at least one value of $\eta$, such that for $n$ sets of measures, all the results are ``success", is given by
\begin{equation}
P_n\leq1-(1-2^{-n})^{2n+1}\,.
\end{equation}
It is clear that $\lim_{n\rightarrow\infty}P_n=0$. This implies, for a sufficiently large $n$, it is impossible that all $n$ sets of measurements result in ``success" for all ratios of $\eta_i=2^i$.

With this strategy, we can determine whether an instance $f$ belongs to MAJSAT .
The time complexity of the entire algorithm is $O(n^4)$, meaning that it runs in polynomial time.

\subsection{LQC algorithm for \pnp}

According to computational complexity theory, we have $\text{P}^{\text{PP}}=\text{P}^{\sharp\text{P}}$~\cite{zoo}. Consequently, the LQC algorithm for PP problems can be adapted to efficiently solve problems in $\sharp \text{P}$ and
${\text P}^{\sharp \text{P}}$. In this section, we briefly discuss the algorithm through a specific example, namely,
the MAX-$k$-IS problem.

The class $\sharp \text{P}$ contains problems where the task is to compute the number of accepting paths in a nondeterministic polynomial-time TM. It is a counting version of the class P, which contains decision problems solvable in polynomial time.  A $\sharp \text{P}$-complete problem is $\sharp$SAT, which is to determine, for a Boolean expression $f(x_1,x_2,\ldots,x_n)$, the number of assignments of Boolean
variables $x_1,x_2,\ldots,x_n$ such that $f=1$. The class ${\text P}^{\sharp \text{P}}$ is defined as the set of problems
that can be deterministically solved in polynomial time with access to a $\sharp \text{P}$-oracle.

We focus on a ${\text P}^{\sharp \text{P}}$ problem called MAX-$k$-IS. For a given graph $G(n,m)$,
there are many ISs. Let us denote
the set of ISs  having $k$  vertices as $k$-IS and its size as $\sharp k$-IS.
For example,  $\sharp 0$-IS is one and $\sharp 1$-IS is $n$. The problem of  MAX-$k$-IS is to determine
which $\sharp k$-IS is the largest. MAX-$k$-IS is evidently a ${\text P}^{\sharp \text{P}}$ problem.
As the $k$-IS is an NP-complete problem, we can query the $\sharp \text{P}$-oracle
for $\sharp 0$-IS, $\sharp 1$-IS, up to $\sharp n$-IS, respectively, and compare them to
determine which  is the largest. We will now show that this problem can be solved by LQC,
using the algorithm for solving PP as shown in Fig.~\ref{t9}.

Regarding the graph $G(n,m)$, similar to the previous section, we employ binary notation:
$x=x_1x_2\ldots x_n$, where $x_j=1$ denotes the selection of the $j$th vertex.
Since the $k$-IS is an NP problem, there exists a Boolean expression $f_k(x_1,x_2,\ldots,x_n)$
such that $f_k=1$ if and only if $x_1x_2\ldots x_n$ forms an independent set containing $k$ vertices.
To solve this problem with the PP algorithm, we formulate an additional Boolean
expression $g_z(x_1,x_2,\ldots,x_n)$, where $1\leq z\leq2^n$.
This expression evaluates to $1$ ($g_z=1$) if the string $x=x_1x_2\ldots x_n$, interpreted as a binary number, is less than $z$.

We next introduce an additional Boolean variable $x_0$, and construct a Boolean expression
involving $n+1$ variables,
\be
F(x_0,x_1,\ldots x_{n}) =(x_0\wedge f_k(x_1,x_2,\ldots,x_n))
\vee(\bar{x}_0\wedge g_z(x_1,x_2,\ldots,x_n)).
\label{Nexpre}
\ee
The expression $F$ is true only when either $f_k$ or $g_z$ is true, and $x_0$ here serves as a switch.

With the Boolean expression (\ref{Nexpre}), we construct a MAJSAT problem:
whether the majority of the assignments for $x_0,x_1,\ldots,x_{n+1}$ satisfy $F=1$. In other words,
we determine whether the following inequality holds:
\begin{equation} \label{cccon}
z+\sharp k{\text -} {\text IS} \geq 2^{n+1}/2=2^n,
\end{equation}
The LQC circuit illustrated in Fig.~\ref{t9}, with $n+1$ work qubits,
can be used to solve this problem. As mentioned above, the time complexity of this algorithm is $O(n^4)$.

The detailed procedure is as follows. For a given $k$, we initiate the process by setting $z=2^{n-1}$, denoted as $z=100\ldots 0$ with $n-1$ zeros,  and formulate the Boolean expression $g_{2^{n-1}}$ to determine whether  $\sharp k$-IS$+2^{n-1}\geq 2^n$ holds. If the result is negative, we keep the first $1$ and change
the first $0$ to $1$, i.e., set $z=1100\ldots 0$; if positive, we  change the first $1$ to $0$ and set $z=0100\ldots 0$. This process is then iterated to determine the subsequent numbers in the binary representation of $z$. The iteration continues until the minimum number $z$ satisfying Eq.~(\ref{cccon}) is obtained, denoted as $z_{\text{MIN}}$. The value of $\sharp k$-IS is then
calculated as $2^n-z_{\text{MIN}}$. The time complexity of this search is $O(n)\cdot O(n^4)=O(n^5)$.

After applying this iterative process for $0$-IS, $1$-IS, $\cdots$, up to $n$-IS,  we have the solution for MAX-$k$-IS.
The total time complexity is $O(n^6)$, which is polynomial. It is important to note once again
that the length of the input for the graph $G(n,m)$ is not $n$ but $n^2$.

The MAX-$k$-IS problem can be readily reformulated as a decision problem. In this formulation, the input takes the form ``$G(n,m),k$", where $1\leq k\leq n$. The algorithm accepts the input if the number of independent sets with exactly $k$ vertices is the largest compared to the number of independent sets with $1,2,\ldots,n$ vertices; otherwise, it rejects the input. Thus, \pnp can be seen as a decision problem, which can be effectively addressed using an LQC algorithm. In any case, this decision can be implemented by an additional oracle, with the additional oracle qubit serving as the ``Y qubit", as indicated in Eq.~(\ref{ZZZZ}).

The above result implies that the class ${\text P}^{\sharp \text{P}}$ is a subset of BLQP.
The entire class PH is defined as: $\Delta_0=\Sigma_0=\Pi_0=\text{P}$;
$\Delta_i=\text{P}^{\Sigma_{i-1}},\Sigma_i=\text{NP}^{\Sigma_{i-1}},\Pi_i=\text{co-NP}^{\Sigma_{i-1}}$.
According to Toda's theorem~\cite{Toda}, $\text{PH}\subseteq{\text P}^{\sharp \text{P}}$.
So, the class PH is a subset of BLQP.


\section{BLQP $=$ \pnp}
Following the proof of BQP$\subseteq$\pnp\cite{Bernstein}, with some modifications, we now proceed to prove
BLQP$\subseteq$\pnp. Consider a polynomially sized LQC algorithm made of a sequence of
logical gates $L_1$, $L_2$, ..., $L_t$. Here $t=p(n)$ with $p(n)$ denoting a polynomial function of the length $n$ of the input.

\subsection{proof of BLQP$\subseteq$\pnp from the perspective of computational problem}

In fact, we can prove that BLQP$\subseteq$\pnp using various methods.
First, suppose there is an algorithm that can obtain a solution encoded by the state $|\text{sol})=|o_1o_2,\ldots,o_n\rangle\otimes|00\ldots 0)$ with probability $P \simeq 1$, i.e.,
\be
e^{\text{i}\theta}|o_1o_2\ldots o_n\rangle\otimes|00\ldots 0)\simeq  L_tL_{t-1}\ldots L_1|x),
\ee
where $|x)=\ket{x}\otimes |00\cdots 0)$ denotes the initial overall state with $n$ qubits and $q(n)$ hybits, where $q(n)$ is at most a polynomial function of $n$, and $\theta$ an arbitrary phase. The state $|00 \ldots 0)$ indicates that all the hybits are in the $|0)$ state (In the context of LQC discussed in Sec.~II, the state becomes unobservable if any hybit is in the $|1)$ state). Thus, the initial state and a meaningful final state encoding the output should have the form $|\ldots)=\ket{\ldots}\otimes |00\cdots 0)$. In contrast, intermediate states and non-meaningful final states do not need to adhere to this constraint.
Following the convention outlined in Sec.~II, the notation $|\thickspace)$ is used for an indefinite inner product state, as the hybits are included in the overall state.
Here, $|o_i\rangle$ represents the state of the $i$th qubit, where $o_i$ can be either $0$ or $1$. Our algorithm for solving the MIS problem is an example of such an algorithm.

We will next demonstrate that this algorithm can be implemented by a polynomial-time deterministic algorithm equipped with a $\sharp \text{P}$ oracle. To achieve this, we need to show how to efficiently compute the following amplitudes on a classical computer equipped with a $\sharp \text{P}$ oracle.,
\be \label{00}
|A(y)|^2=(x|L_1^\dag L_2^\dag\ldots L_t^\dag| y)(y|L_tL_{t-1}\ldots L_1|x),
\ee
where $|y)$ is one of the possible outputs (right for $|\text{sol})$ or wrong for all others).
As any state with $|1)$ is unobservable, we only need consider $|y)$ of the following form
\be
|y)=\ket{y}\otimes |00\ldots 0)\,.
\ee

The amplitude $A$ can be decomposed as:
\be \label{zz}
|A(y)|^2=\sum_{z_1,z_2,z_{2t-2}} (x|L_1^\dag|z_{2t-2})\ldots (z_t|L_t^\dag|y)( y|L_t|z_{t-1})\ldots ( z_1|L_1|x)\,,
\ee
where the summation is over all possible $z_i$'s. It is important to note that in the computational basis, we still have $\sum_i|z_{i})( z_{i}|=1$.

Computing any of the terms in the summation (\ref{zz})
is a polynomially sized task as $t=p(n)$ is a polynomial function of $n$ and each $L_i$, in its matrix form,
has only finite number of non-zero elements. This allows us introduce a language, denoted
by $+r$-DTM, defined by a number $c$. The input is $\{|y),|z_1),\ldots |z_{2t-2}),|x),L_1,\ldots,L_t,k\}$
with $1\leq k\leq 2^{2^c}$ being an integer. It is clear that the total length
of the input is a polynomial function of $n$. The task is to deterministically calculate
\be   \label{haowan}
a=(x|L_1^\dag|z_{2t-2})\ldots (z_t|L_t^\dag|y)( y|L_t|z_{t-1})\ldots ( z_1|L_1|x),
\ee
and then decide whether $\text{Re}(a)>0$ and $k<\text{Re}(a)$. If this is true,
the output of the $+r$-DTM is $M=1$ (accept);
otherwise, $M=0$ (reject). This is a P problem. Following the proof of  BQP$\subseteq$\pnp\cite{Bernstein},
by counting how many times $M=1$, one can efficiently compute the positive real part of the following summation,
\be
|A_s|^2=\sum_{y\in s}|A(y)|^2\,,
\ee
where $s$ is a subset of all $2^{n}$ possible values of $y$ associated with the $n$ qubits. For example, $s=\{00\cdots 01, 10\cdots 01\}$
and $s=\{0,1\}^{n}$. Similarly, we can compute efficiently the negative real part and finally $|A_s|^2$.

At this stage, concerns may arise regarding the precision of $a$ in Eq.~(\ref{haowan}), which, in the algorithm described above, is of the order $O(1)$. However, this is not problematic because: (i) the precision can be easily adjusted, and (ii) for the powerful LQC algorithm, the contribution to the amplitude is exponentially significant relative to the input length $n$. Therefore, a precision of order $O(1)$ suffices. The parameter $c$ is chosen based on the specific problem instance, specifically the maximum value $k$ in the TM's input, where $2^{2^c}$ exceeds the maximum output $|a|$ of the TM. For sufficiently large $n$, we typically set $c = n$, ensuring that $2^{2^n}$ is significantly larger than both the maximum output $|a|$ of the TM and consequently the total amplitude $|A_s|^2$, which typically scales as $\exp(n)$ for the LQC algorithm.

We first compute $|A_s|^2$ for $s=\{0,1\}^n$ and denote the result as $W_0$. This is the total amplitude,
which is one for BQP. However, for BLQP, this is usually a large number. Next, we represent an element in $s$ that denote the qubits' states as an $n$-digit binary number. We then compute $|A_s|^2$
for $s=\{y|y<2^{n-1}\}$, and denote it as $W_1$. If $W_1/W_0\lesssim \varepsilon$, where $\varepsilon$
is a small number decreasing exponentially with $n$ and the symbol ``$\lesssim$" means ``less than" and ``almost equal to", then we compute $|A_s|^2$ for $s=\{y|2^{n-1}\le y <3\times 2^{n-1}\}$; otherwise, we compute $|A_s|^2$
for $s=\{y|y<2^{n-2}\}$. In either case, we denote the result as $W_2$. If $W_2/W_0\lesssim 1$, then we cut the current interval into half and compute $|A_s|^2$
for one of the halves; otherwise, we cut the other interval into half and compute $|A_s|^2$
for one of the halves. In either case, we denote the result as $W_3$, and so on. We continue this for $n$ steps, and
will eventually arrive at a $y$ for which $|A(y)|^2\sim W_0$. This $y$ is the solution $|o_1o_2\ldots o_n\rangle\otimes|00\ldots 0)$.

Clearly, this branching algorithm can be extended to cases with polynomially many solutions:
\be
c_1 |\text{sol}_1)+c_2 |\text{sol}_2)+\ldots  \simeq  L_tL_{t-1}\ldots L_1|x),
\ee
where $|c_1|\simeq |c_2|\simeq\ldots$. In this case, we only need to perform polynomial branching to find all the solutions.

So far, we have proven that if an Lorentz quantum circuit can generate a specific state or a superposition of polynomially many states that satisfy the conditions given by the problem, by designing the sequence of gates $L_1, L_2, \ldots, L_t$, then it can be implemented by a polynomial-time deterministic algorithm with a $\sharp \text{P}$ oracle.


\subsection{proof of BLQP$\subseteq$\pnp by decision problem}

Since in computational complexity theory any problem is defined as a decision problem, we now examine the decision problem to provide a more formal proof that BLQP$\subseteq$\pnp.

For a decision problem where the criterion for an accepting state is encoded in a ``Y qubit" in the context of BLQP defined in Sec.~II (refer to Eq.~(\ref{ZZZZ})), it is convenient to introduce a projection operator $P_{\text{yes}}$,
\be \label{YES}
P_{\text{yes}}=\left[\sum_{j}|j_{Q_1}\rangle\langle j_{Q_1}|\otimes\ldots\otimes\sum_{j}|j_{Q_{n-1}}\rangle\langle j_{Q_{n-1}}|\right]\otimes |1_Y\rangle\langle 1_Y|  \otimes \left[ |0_{H_1}) ( 0_{H_1}|\otimes\ldots\otimes |0_{H_{q(n)}}) (0_{H_{q(n)}}|\right],
\ee
where $n$ is the number of qubits in the current LQC and $q(n)$ the number of hybits, $|\ldots_Y\rangle$​​ denotes the state of the ``Y qubit", $|\ldots_{Qi}\rangle$​​ denotes the state of the $i$th qubit and $|\ldots_{Hi})$​​ denotes the state of the $i$th hybit.

With the projection operator $P_{\text{yes}}$, we have,
\be \label{01}
|c_{\text{yes}}|^2=(x| L_1^\dag L_2^\dag \ldots L_t^\dag|P_{\text{yes}}|L_tL_{t-1}\ldots L_1|x),
\ee
where $|c_{\text{yes}}|^2$ is precisely as shown in Eq.~(\ref{ZZZZ}).

In the discussion regarding the relationship between BLQP and \pnp​​, the amplitude $|c_{\text{yes}}|^2$ in Eq.~(\ref{01}) is decomposed as follows:
\ba \label{1} \nonumber
|c_{\text{yes}}|^2&=&\sum_{z_1,z_2,\ldots,z_{2t}} (x|L_1^\dag|z_{2t})( z_{2t}|L_2^\dag|z_{2t-1})\ldots (z_{t+2}|L_t^\dag|z_{t+1})
  ( z_{t+1}| P_{\text{yes}}|z_{t})\\ &&( z_{t}|L_t|z_{t-1})   \ldots ( z_{2}|L_2|z_{1})( z_{1}|L_1|x),
\ea
where each $|z_i)$ represents a complete orthonormal basis vector, specified within the computational basis.

We introduce a language within the class P, determined by $+r$-DTM, defined by a number $c$.
The inputs are in the form ($|z_1),\ldots |z_{2t}),|x\rangle,L_1,\ldots,L_t,k$) where $1\leq k\leq 2^{2^c}$ is an integer for each input, and the total length of the input is a linear function of $p(n)$, thus remaining polynomial in $n$, since $p(n)$ is a polynomial function of $n$.

Given an input, we deterministically calculate:
\be
a=(x|L_1^\dag|z_{2t})( z_{2t}|L_2^\dag|z_{2t-1})\ldots (z_{t+2}|L_t^\dag|z_{t+1})
  ( z_{t+1}| P_{\text{yes}}|z_{t})( z_{t}|L_t|z_{t-1})   \ldots ( z_{2}|L_2|z_{1})( z_{1}|L_1|x).
\ee
If $\text{Re}(a)>0$ and $k<\text{Re}(a)$, the $+r$-DTM outputs $M=1$ (accept); otherwise, it outputs $M=0$ (reject). Each gate $L_i$ is associated with only a few bits. Although $P_{\text{yes}}$​​, as shown in Eq.~(\ref{YES}), contains an exponential number of elements in the computational basis, calculating $( z_{t+1}| P_{\text{yes}}|z_{t})$ is straightforward and thus takes only a single step. Therefore, determining the output of the $+r$-DTM as $M=1$ or $0$ is a problem within class P when considering the length $n$ of the input to the LQC.

If we have access to a $\sharp \text{P}$-oracle, which can inform us of the number of accepted paths, with each path representing a problem within class P, we can query this oracle to determine how many times $M=1$ out of all possible inputs of ($|z_1),\ldots|z_{2t}),|x),L_1,\ldots,L_t,k$). This number essentially represents the real positive portion of the sum:
\be
\sum_{z_1,z_2,\ldots,z_{2t}} (x|L_1^\dag|z_{2t})( z_{2t}|L_2^\dag|z_{2t-1})\ldots (z_{t+2}|L_t^\dag|z_{t+1})
  ( z_{t+1}| P_{\text{yes}}|z_{t})( z_{t}|L_t|z_{t-1})   \ldots ( z_{2}|L_2|z_{1})( z_{1}|L_1|x).
\ee
Thus, by utilizing a $\sharp \text{P}$-oracle, we can directly obtain the positive real part of the final amplitude of a Lorentz quantum circuit. Similarly, we define $-r$-DTM and finally obtain $|c_{\text{yes}}|^2$.


To accurately compute the acceptance probability in the LQC algorithm, it is crucial to determine the total effective amplitude. This is not simply the identity ``1", but rather the total ``observable probability", given by $|c_{\text{yes}}|^2+|c_{\text{no}}|^2$. This refers to the observable portion of the final state where all hybits are in the state $|0)$ (see Eq.~(\ref{ZZZZ}) for details). This quantity corresponds to $W_0$ in the branching algorithm described earlier. Alternatively, it can be obtained similarly to $|c_{\text{yes}}|^2$ using a slightly different projection operator $P_{\text{yn}}$ instead of $P_{\text{yes}}$:
\be \label{YES2}
P_{\text{yn}}=\left[\sum_{j}|j_{Q_1}\rangle\langle j_{Q_1}|\otimes\ldots\otimes\sum_{j}|j_{Q_{n-1}}\rangle\langle j_{Q_{n-1}}|\right]\otimes \sum_{j} |j_Y\rangle\langle j_Y|  \otimes \left[ |0_{H_1}) ( 0_{H_1}|\otimes\ldots\otimes |0_{H_{q(n)}}) (0_{H_{q(n)}}|\right].
\ee
The total amplitude can then be expressed as:
\ba \label{122} \nonumber
|c_{\text{yes}}|^2+|c_{\text{no}}|^2&=&\sum_{z_1,z_2,\ldots,z_{2t}} (x|L_1^\dag|z_{2t})( z_{2t}|L_2^\dag|z_{2t-1})\ldots (z_{t+2}|L_t^\dag|z_{t+1}) \\
 && ( z_{t+1}| P_{\text{yn}}|z_{t})( z_{t}|L_t|z_{t-1})   \ldots ( z_{2}|L_2|z_{1})( z_{1}|L_1|x),
\ea
This quantity can be obtained using the same method by querying the $\sharp P$-oracle.

The probability of accepting the input in the LQC circuit is then calculated as $\frac{|c_{\text{yes}}|^2}{|c_{\text{yes}}|^2+|c_{\text{no}}|^2}$​​. The \pnp algorithm outputs $1$ (accept) if $\frac{|c_{\text{yes}}|^2}{|c_{\text{yes}}|^2+|c_{\text{no}}|^2}\geq2/3$, and $0$ (reject) if $\frac{|c_{\text{yes}}|^2}{|c_{\text{yes}}|^2+|c_{\text{no}}|^2}\leq1/3$. Thus, we have shown that BLQP$\subseteq$\pnp, meaning that if an input can be decided by some BLQP algorithm, it can also be decided by a \pnp algorithm.

Given that we previously established \pnp$\subseteq$BLQP, we conclude that BLQP$=$\pnp.

We can also generalize the proof for BQP $\subseteq$ PSPACE ~\cite{Arora,Nielson} to prove
BLQP $\subseteq$ PSPACE. This generalization is rather trivial and straightforward as
the proof for BQP $\subseteq$ PSPACE does not require unitarity for the gate transformation.
We have not found an efficient LQC algorithm for the problem of
quantified Boolean formulas (QBF), recognized as PSPACE-complete~\cite{zoo,Papadimitriou}.

For quantum computing with postselection, it was established that
${\text {PostBQP}}={\text {BQP}}^{{\text {PostBQP}}}_{\parallel,\text{classical}}$~\cite{Post}.
Similarly, we have ${\text {BLQP}}={\text {BQP}}^{{\text {BLQP}}}_{\parallel,\text{classical}}$.
As BLQP$=$\pnp,  we can directly conclude that
\be
\text{\pnp}={\text {BQP}}^{{\text {\pnp}}}_{\parallel,\text{classical}}.
\ee
More interesting results are expected for the class \pnp in light of  the new perspective provided by
BLQP$=$\pnp.

\section{Comparison between LQC and quantum computing with postselection}

As far as we know, the term ``postselection"  has several meanings.
Quite often it refers to a method of selectively choosing specific outcomes after many rounds of quantum measurements~\cite{peres,Aharonov}. The postselection that we are discussing here was introduced
by Aaronson as ``the power of discarding all runs of a computation in which a
given event does not occur"~\cite{Post}. In other words, it is a capability to force specific outcomes
in a single run of quantum measurement, which is beyond quantum mechanics.
A quantum computer with this ability of postselection has been found to be very powerful,  and
the corresponding computational complexity class PostBQP was shown to be equivalent to PP~\cite{Post}.
Below we briefly review this concept and then discuss the relationship between PostBQP and BLQP.

\subsection{Simulation of postselection by LQC}
The postselection introduced in Ref.~\cite{Post} is the ability
to efficiently collapse a quantum state given by
\begin{equation} \label{post1}
|\Psi\rangle=\sum_i c_i|\psi_i\rangle=\sum_{j\in\text{yes}}c_j|\psi_j\rangle+\sum_{k\in\text{no}}c_k|\psi_k\rangle,
\end{equation}
to the following target state,
\begin{equation}  \label{post2}
|\Psi_{\text{yes}}\rangle=\frac{1}{\sum_{j\in\text{yes}}|c_j|^2}\sum_{j\in\text{yes}}c_j|\psi_j\rangle.
\end{equation}
Here $|\Psi\rangle$ represents a general quantum state, and $|\psi_i\rangle$s are basis states that are
categorized into  `yes' and `no' according to a given problem.

Basically, the postselection consists of two operations. The first one behaves like an oracle,  marking
each state as 'yes' or 'no'. The second is quantum measurement with the  ability to collapse to only yes states.
Both of the operations can be simulated by LQC: the first one with an oracle qubit and
the second one with Lorentz transformations on a hybit.  And
the  states $|\Psi\rangle$ and $|\Psi_{\text{yes}}\rangle$ are stored in $n$ work qubits.

The specific process unfolds as follows.
The initial state is prepared as
\begin{equation} \label{post11}
|\Psi_i)=|0_o\rangle\otimes|\Psi\rangle\otimes|0).
\end{equation}
Here $|0_o\rangle$ is the state of the oracle qubit, and $|0)$ is for the hybit, which can only undergo
Lorentz transformation in the space spanned by $|0)$ and $|1)$.
After the oracle operation, which is described in Eq.~(\ref{Oracle}) and illustrated in
the small box marked ``oracle" in Fig.~\ref{t7}, the state of the system becomes
\begin{align}
|\Psi_o)=|1_o\rangle\otimes\sum_{j\in\text{yes}}c_j|\psi_j\rangle\otimes|0)
+|0_o\rangle\otimes\sum_{k\in\text{no}}c_k|\psi_k\rangle\otimes|0)\,,
\label{post12}
\end{align}
where the 'yes' states and 'no' states are marked out with the oracle qubit.
Within the oracle, whether a given state
$|\psi_i\rangle$ belongs to 'yes' or 'no' can be verified within polynomial time.
This oracle operation  can be implemented with a conventional quantum computer and the
states in superposition are checked in parallel.

We then use a manipulation that is unique in LQC.
It is the CV gate shown in Fig.~\ref{t5}, with the oracle qubit  as the control and the hybit as the target. When
the oracle qubit is in the state of $\ket{1}$, a Lorentz transformation $V$ in Eq.~(\ref{Ltransformation})
is applied to the hybit. After applying the CV gate
 $r$ times, we have
 \begin{align}
|\Psi_V)=&|1_o\rangle\otimes\sum_{j\in\text{yes}}c_j\cosh(r\chi)|\psi_j\rangle\otimes|0)
+|0_o\rangle\otimes\sum_{k\in\text{no}}c_k|\psi_k\rangle\otimes|0)\nonumber  \\
+&|1_o\rangle\otimes\sum_{j\in\text{yes}}c_j\text{i}\sinh(r\chi)|\psi_j\rangle\otimes|1)\,.
\label{post13}
\end{align}
Because the state $|1)$ for the hybit is unobservable, the resulting state is equivalent to
\begin{align}
|\Psi_V)=|1_o\rangle\otimes \cosh(r\chi)\sum_{j\in\text{yes}}c_j|\psi_j\rangle\otimes|0)
+|0_o\rangle\otimes\sum_{k\in\text{no}}c_k|\psi_k\rangle\otimes|0)\,.
\label{post14}
\end{align}
where the normalization constant is omitted. If
the repetition time is $r\approx\frac{1}{\chi}\ln 2^n\sim O(n)$, the amplitude for the 'yes' states is exponentially
larger. If  the oracle bit is measured, it is almost certain
to find it in the state of $\ket{1_o}$. Thus, after the measurement, we are almost certain to
find the system in the following state
\begin{equation} \label{post15}
|\Psi_f)=|1_o\rangle\otimes\sum_{j\in\text{yes}}c_j|\psi_j\rangle\otimes|0).
\end{equation}
The postselection is accomplished.

It is apparent that even when the absolute values of $|c_j|$s are very different, e.g.,
some $|c_j|$s are exponentially smaller than the others, we can still obtain the states (\ref{post15}) in polynomial time in $n$.

\subsection{Super-postselection by LQC}
In Sec.~\ref{sec:mis}, we have discussed an operation called $Q$, whose LQC circuit is shown in
Fig.~\ref{QC}. The $Q$ operation  is capable of counting the number of qubits in the state $\ket{1}$ within $\ket{\psi_i}$,
and then uses this count to appropriately amplify the amplitude. By repeating $Q$ sufficient
number of times, one can select the basis states that
have the largest number of qubits in $\ket{1}$.

Consider, for example,  a superposition state given by
\begin{equation} \label{post21}
|\Phi_0)=\big[\ket{1000}
+\ket{0110}\big]\otimes\ket{1_o}\otimes |0),
\end{equation}
which is not normalized because the normalization is not important.
We apply  $Q$ with four work qubits to this state $r$ times,
and the state becomes
\begin{align} \label{post22}
|\Phi_r)
=&\big[\cosh(r\chi)\ket{1000}
+\cosh(2r\chi)\ket{0110}\big]\otimes\ket{1_o}\otimes |0)\nonumber\\
+&\text{i}\big[\sinh(r\chi)\ket{1000}
+\sinh(2r\chi)\ket{0110}\big]\otimes\ket{1_o}\otimes |1)\,.
\end{align}
For $\chi=4\ln(\sqrt{2}+1)$, when $r=4$, the ratio
between the coefficient before $\ket{1000}$ and the coefficient before $\ket{0110}$ is over $10^6$.
Because $|1)$ is unobservable,
the above state can be regarded approximately as
\be
|\Phi_r)\approx \ket{0110}\otimes\ket{1_o}\otimes |0)\,.
\ee
The state $\ket{0110}$ is selected.  However, let us start with
a different superposition state
\begin{equation} \label{post121}
|\Phi_0)=\big[\ket{1110}
+\ket{0110}\big]\otimes\ket{1_o}\otimes |0)\,.
\end{equation}
This time we do the same $Q$ operation four times.
What is selected at the end is $\ket{1110}$ instead of $\ket{0110}$
because $\ket{1110}$ has more ones than $\ket{0110}$.

From the above example, it is clear that the selection
achieved by repeated $Q$ operations is relative. This
is in stark contrast to postselection, which is done according to
a preset criterion.
It is precisely due to this special selection capability of LQC
that we are able to solve the MIS problem  in polynomial time with the circuit shown in Fig.~\ref{t7}.
We call the selection by $Q$ super-postselection
just to distinguish it from Aaronson's postselection introduced in Ref.~\cite{Post}.

The above comparison shows that LQC can efficiently solve any problem that can be efficiently
solved by a quantum computer with postselection. However, the reverse is not necessarily true
due to the super-postselection capability of LQC. This implies that the complexity class
PostBQP is a subset of BLQP, but it may not be a strict subset. The reason for this is that
a problem that LQC solves with super-postselection may be  solved efficiently by postselection using a different strategy.

\section{Summary}

In summary, we have demonstrated the superior power of
the Lorentz quantum computer (LQC) through concrete examples.
These results show that its computational complexity class
BLQP  (bounded-error Lorentz quantum polynomial-time)
is equivalent to \pnp.
In comparison, it is not even clear whether the complexity class BQP associated with the conventional quantum computer contains NP or not. Our work will likely motivate further study into LQC to better understand its capabilities.

This work also reveals a fascinating relation between computational power and physics. In Ref.~\cite{Post}, it is argued that quantum mechanics is an island in the ``theoryspace".
LQC appears to put an intriguing twist on this claim. On one hand, Lorentz quantum mechanics seems drastically different from quantum mechanics by having unobservable states while living in an indefinite inner product space with complex Lorentz transformations~\cite{Dirac,Pauli}.
On the other hand, the Bogoliubov exications, quasi-particles of bosonic many-body systems, do behave approximately  like a Lorentz quantum mechanical system~\cite{wunjp}.

\vskip 10pt
\appendix
			\renewcommand\theequation{\thesection\arabic{equation}} 

\section{State of the oracle qubit, the auxiliary qubit, and the hybit}
After the step (iv) of the algorithm solving MAJSAT, we have (dropping the state for work qubits $|11\ldots1\rangle$ )
\be
|\psi_{\rm iv})=\ket{\phi_o}\otimes |0)
\otimes \ket{\varphi_{\beta/\alpha}}
\ee
where $\ket{\phi_o}=a_s\ket{0_o}+b_s\ket{1_o}$
with
\be
a_s=\frac{N-s}{\sqrt{(N-s)^2+s^2}}\,,~~~b_s=\frac{s}{\sqrt{(N-s)^2+s^2}}\,.
\ee
After applying the controlled-Hadamard gate at the step (v), we have
\begin{align}
|\psi_{\rm v})&= \alpha \ket{\phi_o}\otimes |0)
\otimes \ket{0}+
\frac{\beta}{\sqrt{2}}\big[(a_s+b_s)\ket{0_o}+(a_s-b_s)\ket{1_o}\big]\otimes |0)\otimes \ket{1}\nonumber\\
&=\ket{0_o}\otimes |0) \otimes\Big[\alpha a_s\ket{0}+\frac{\beta(a_s+b_s)}{\sqrt{2}}\ket{1}\Big]+
\ket{1_o}\otimes |0) \otimes\Big[\alpha b_s\ket{0}+\frac{\beta(a_s-b_s)}{\sqrt{2}}\ket{1}\Big]
\end{align}
After the application of CV gate $r'$ time at the step (vi) and the omission
of terms that are either unobservable or exponentially small,  this state becomes
\be
|\psi_{\rm vi})\approx
\ket{1_o}\otimes |0) \otimes\Big[\alpha b_s\ket{0}+\frac{\beta(a_s-b_s)}{\sqrt{2}}\ket{1}\Big]\,.
\ee
The exact probability of having this state is
\be
P_{\beta/\alpha}=\frac{\cosh^2(2r\chi)}{2^n-1+\cosh^2(2r\chi)}
\cdot \frac{\cosh^2(r'\chi)(\alpha^2s^2+\beta^2(2^n-2s)^2/2)} {\cosh^2(r'\chi)(\alpha^2s^2+\beta^2(2^n-2s)^2)/2+\alpha^2(2^n-s)^2+\beta^2 2^{2n-1}}\,,
\ee
where $\chi=2\ln(\sqrt{2}+1)$.  We have $P_{\beta/\alpha}\approx 1$
when $r\approx r'\approx \ln N/\chi\propto n$.

\section{Possible values of $\eta=\beta/\alpha$}
The probability of obtaining $-1$ when the auxiliary qubit state $\varphi_{\eta}$ is measured along the $x$-direction is
\begin{align}
P_-&=|\langle-|\varphi_{\eta}\rangle|^2\nonumber\\
&=\frac{|s-\eta\sqrt{1/2}(2^n-2s)|^2}{2\big[s^2+\eta^2(2^n-2s)^2/2\big]}\nonumber\\
&=\frac{s^2+\eta^2(2^{n}-s)^2/2-\sqrt{2}\eta s(2^n-2s)}{2\big[s^2+\eta^2(2^n-2s)^2/2\big]}\nonumber\\
&=\frac12-\frac{\sqrt{2}\eta s(2^n-2s)}{2s^2+\eta^2(2^n-2s)^2}
\end{align}
When $s>2^{n-1}$, the second term on the right hand side is positive and we re-write it as
\be
P_--\frac12=\frac{\sqrt{2}\eta s(2s-2^n)}{2s^2+\eta^2(2s-2^n)^2}\,.
\ee
It can be shown, for a given probability $0<\delta_p<1/2$, when
\be
\frac{(1-\sqrt{1-4\delta_p^2})s}{\sqrt{2}\delta_p(2s-2^n)}\leq \eta \leq \frac{(1+\sqrt{1-4\delta_p^2})s}{\sqrt{2}\delta_p(2s-2^n)}\,,
\ee
we have
\be
\label{diff}
P_--\frac12\geq \delta_p\,.
\ee
Let $s=2^{n-1}+\delta s$ and we have
\be
\frac{(1-\sqrt{1-4\delta_p^2})s}{2\sqrt{2}\delta_p\,\delta s}\leq \eta \leq \frac{(1+\sqrt{1-4\delta_p^2})s}{2\sqrt{2}\delta_p\,\delta s}\,,
\ee
When $\delta s=2^{n-1}$ we have the ratio $s/\delta s=2$, which is the smallest.
This shows that to satisfy the inequality (\ref{diff}) for a given $\delta_p$ for all possible values of $s$, we must have
\be
\eta_m-\eta_s=\frac{2\sqrt{1-4\delta_p^2}}{\sqrt{2}\delta_p}\,,
\ee
where $\eta_m$ and $\eta_s$ are the largest and smallest values
of $\eta$. Since $\eta$ takes only discrete values
in the form of $2^{j}$ with $j\in [-n,n]$, the number of $\eta$
satisfying   the inequality (\ref{diff}) increases with
smaller $\delta_p$.

\section{Boolean expression for the function $g_z$} \label{Appendix_A}
The Boolean formula $g_z(x_1,x_2,\cdots,x_n)$ used in the main text is defined as
$g_z=1$ if the string $x=x_1x_2\cdots x_n$, interpreted as a binary number, is less than $z$.
When $x\geq z$, we have $g_z=0$.

We use an example to show how to construct $g_z(x_1,x_2,\cdots,x_n)$. Assume
that $n=8$ and $z=10101001$, we then have
\be
g_z=\bar{x}_1\vee(\bar{x}_2\wedge\bar{x}_3)\vee(\bar{x}_2\wedge\bar{x}_4\wedge\bar{x}_5)\vee(\bar{x}_2\wedge\bar{x}_4
\wedge\bar{x}_6\wedge\bar{x}_7\wedge\bar{x}_8)\,.
\ee
The length of the Boolean expression thus constructed is always less than $n^2$. This construction method is applicable to different values of $N$.


%

\end{CJK*}

\end{document}